\documentclass[prd,amsfonts,onecolumn,superscriptaddress,aps,nofootinbib,11pt]{revtex4}

\pdfoutput=1
\usepackage[utf8]{inputenc}
\usepackage{amsmath,amssymb,mathrsfs}
\usepackage{graphicx}
\usepackage{enumerate}
\usepackage{multirow}
\usepackage[usenames,dvipsnames]{xcolor} 
\usepackage{hyperref}
\hypersetup{colorlinks=true,
		breaklinks=true,
    linkcolor=Blue,          % color of internal links
    citecolor=Plum,        % color of links to bibliography
    urlcolor=YellowOrange           % color of external links
}
\usepackage[]{units}
\usepackage{ulem} 

\providecommand{\ums}[2][1]{\ml{\tfrac{#1}{#2}}}

\providecommand{\ZZ}{\mathbb{Z}}
\providecommand{\ml}[1]{\mbox{\large $#1$}}

\providecommand{\to}{\rightarrow}

\providecommand{\bs}[1]{\boldsymbol{#1}}

\usepackage{tabu}
\providecommand{\be}{ \begin{equation} }
\providecommand{\ee}{\end{equation}}
\providecommand{\bea}{\begin{eqnarray}}
\providecommand{\eea}{\end{eqnarray}}
\providecommand{\nn}{\nonumber}

\providecommand{\eq}[1]{\begin{equation} #1 \end{equation}}

\providecommand{\ums}[2][1]{\ml{\tfrac{#1}{#2}}}

\providecommand{\ZZ}{\mathbb{Z}}
\providecommand{\ml}[1]{\mbox{\large $#1$}}

\providecommand{\to}{\rightarrow}

\begin{document}
%%%%%%%%%%%%%%%%%%%%%%%%%%%%%%%%%%%%%%%%%%%%%%%%%
\title{Fermion Mass Hierarchy and Double Seesaw Mechanism\\ in a 3-3-1 Model with an Axion}

\author{A.~G.~Dias}%
\email{alex.dias@ufabc.edu.br}
\affiliation{Centro de Ci\^encias Naturais e Humanas, Universidade Federal do ABC, Santo Andr\'e-SP, Brasil}

\author{J.~Leite}%
\email{julio.leite@ufabc.edu.br} 
\affiliation{Centro de Ci\^encias Naturais e Humanas, Universidade Federal do ABC, Santo Andr\'e-SP, Brasil}

\author{D.~D.~Lopes}%
\email{lopes.diego@ufabc.edu.br} 
\affiliation{Centro de Ci\^encias Naturais e Humanas, Universidade Federal do ABC, Santo Andr\'e-SP, Brasil}

\author{C.~C.~Nishi}%
\email{celso.nishi@ufabc.edu.br}
\affiliation{Centro de Matem\'atica, Computa\c{c}\~ao e Cogni\c{c}\~ao, Universidade Federal do ABC,  Santo Andr\'e-SP, Brasil}

\date{\today}

%%%%%%%%%%% ABSTRACT %%%%%%%%%%%%%%%%%%%%%%%%%%%%%%%%
\begin{abstract}
We present  a model based on the $SU(3)_C\otimes SU(3)_L\otimes U(1)_X$ gauge symmetry that relates 
the mass hierarchy of the fermions with the solution to the strong CP problem through the $U(1)_{PQ}$ Peccei-Quinn symmetry. 
This last symmetry arises accidentally with the imposition of a discrete $\ZZ_9$ symmetry, 
which also secludes the different scales in the double seesaw mechanism taking place in the neutrino sector.
The symmetry breakdown is performed by three scalar triplets plus a scalar singlet hosting an axion field, whose particle excitation can be a component of dark matter. We show a mechanism where a small effective vev is generated for a scalar triplet which is supposed to have a  bare mass above the energy scale where the $SU(3)_L\otimes U(1)_X$ symmetry is broken. Combined with the energy scale in which the $U(1)_{PQ}$ is broken, such a mechanism gives rise to a natural hierarchy to the fermions. Beyond the Standard Model particle content, the model predicts an invisible axion, $a$, three 
GeV neutrinos, $N_{iL}$, plus several new particles at the TeV scale which are: five vector bosons, $U^\pm$, $V^0$, $V^{0\dagger}$, and $Z^\prime$; one up-type $U$, and two down-type $D_a$ quarks; and at least a CP-even, $H_1$, plus non-hermitian neutral, $\phi^0$,  $\phi^{0\dagger}$, scalar bosons. The model may be tested by looking for the possible production of such particles at the LHC.
\end{abstract}

%%%%%%%%%%%%%%%%%%%%%%%%%%%%%%%%%%%%%%%%%%%%%%%%%
\maketitle
%%%%%%%%%%%%%%%%%%%%%%%%%%%%%%%%%%%%%%%%%%%%%%%%%
\newpage
\section{Introduction}

Despite the great experimental success, the Standard Model (SM) of particle physics leaves unanswered many pressing questions. Some of them are of crucial importance for our understanding of the universe, such as the mechanism behind neutrino masses, the solution to the strong CP problem, and the essence of dark matter. Other open questions, although sometimes treated with less interest, are very intriguing from a theoretical viewpoint and are worth exploring. For example, the reason why there is a strong mass hierarchy among the different fermion families and why the number of families in nature turns out to be exactly three. A common feature that connects these problems is the fact that they all seem to call for physics beyond the SM.

In this paper we present a model based on the $SU(3)_C\otimes SU(3)_L\otimes U(1)_X$ symmetry that belongs to the known class of 3-3-1 models for which the number of fermion generations is not arbitrary but follows from the requirement that all gauge anomalies must cancel~\cite{Singer:1980sw,Montero:1992jk,Foot:1994ym,Hoang:1995vq,Pisano:1991ee,Frampton:1992wt}. Our construction is motivated by two interrelated issues. The first is the strong mass hierarchy among the fermions,
including the even larger mass gap to the neutrinos. We explain this last aspect with the double seesaw mechanism~\cite{Mohapatra:1986aw,Mohapatra:1986bd,Barr:2003nn}. The second issue is the strong CP problem which we solve 
through the Peccei-Quinn (PQ) mechanism~\cite{Peccei:1977hh,Weinberg:1977ma,Wilczek:1977pj}, resulting in an invisible axion that can play a role of dark matter~\cite{Kim:1979if,Shifman:1979if,Dine:1981rt,Zhitnitsky:1980tq}.\footnote{Alternative ways of implementing a PQ symmetry in 3-3-1 models can be found in Refs. \cite{Pal:1994ba, Dias:2003iq, Montero:2017yvy}. }

In our construction, the fermion mass hierarchies and the $U(1)_{PQ}$ Peccei-Quinn symmetry will arise, in the latter case accidentally, from the 3-3-1 gauge structure and an additional $\ZZ_9$ symmetry.
Some of the mass hierarchies will reflect the hierarchy among the different scales in the spontaneous breaking of the gauge group. More specifically, the generation of fermion masses in our model is such that both up-type and down-type quarks present natural mass hierarchies between their third and first two generations. In order to obtain charged lepton masses in agreement with experimental data, the model requires much less suppressed Yukawa couplings than those in the SM. Finally, when it comes to neutrino masses, the double seesaw mechanism is easily implemented after extending the fermion content of Refs.~\cite{Singer:1980sw,Montero:1992jk,Foot:1994ym,Hoang:1995vq} by including singlet neutrino fields.

The spontaneous symmetry breaking is realized by three scalar triplets of $SU(3)_L$ plus a scalar singlet getting a vacuum expectation value (vev). We develop a scheme in which one of the scalar triplets is assumed to have a bare mass above the scale $w$ where the $SU(3)_L\otimes U(1)_X$ is broken to  $SU(2)_L\otimes U(1)_Y$, so that it can be integrated out leading to a low energy effective 3-3-1 model with two scalar triplets. We show that, along with the vev of the scalar singlet breaking the $U(1)_{PQ}$ symmetry, this furnishes a consistent hierarchical fermion mass spectrum.  

In addition to the Standard Model particle content, the model predicts five vector bosons: a singly charged $U^\pm$; a neutral non-hermitian pair $V^{0}$, $V^{0\dagger}$; and a real  $Z^\prime$. The  masses of these vector bosons are expected to be  at the TeV scale, according to our scheme for generating hierarchical masses to the fermions.  Such scheme also leads to the specific prediction for the mass difference of the squared masses of the  $V^{0}$ and $U^\pm$ as being essentially equal to the $W^\pm$ vector boson squared mass, i.e., $M^{2}_{V}-M^2_{U}\simeq M^2_{W}$.  Besides the Higgs boson with mass $m_h\approx 125$ GeV, the scalar particle spectrum up to the TeV scale is composed by a light axion -- the pseudo-Nambu-Goldstone boson of the $U(1)_{PQ}$ symmetry broken at a very high energy scale around $10^{10} \text{ GeV}$  -- plus a CP-even and a non-hermitian neutral fields, $H_1$ and $\phi^0$, both with masses around the TeV scale. The remaining scalars, among which there are two charged fields, have masses well above the TeV scale and outside of the direct LHC reach.

The question of finding an explanation to the hierarchy of fermion masses in a different version of the 3-3-1 model with a minimal scalar sector was first treated in~\cite{Barreto:2017xix}. Our novel contribution to this quest is to demonstrate that the scales required to generate hierarchical fermion masses can be identified in a non-trivial way with those breaking the $SU(3)_L\otimes U(1)_X$ and $U(1)_{PQ}$ symmetries in our 3-3-1 model. 

Other recent studies have also tackled the question of fermion mass hierarchy in 3-3-1 models~\cite{Hernandez:2013hea, Hernandez:2015tna, Hernandez:2016eod, CarcamoHernandez:2017kra, CarcamoHernandez:2018iel, CarcamoHernandez:2017cwi}, but from another perspective. In such works, the observed mass hierarchies follow from the imposition of different discrete flavor symmetries alongside several new scalar fields. In our case, however, we keep the scalar sector as minimal as possible by adding only one scalar singlet and, in addition to a discrete symmetry, our model features an accidental $U(1)_{PQ}$ symmetry.

We organize this work as follows. In the next section we specify the model through its field representation content, the main aspects of the symmetry breaking, and the gauge bosons mass spectrum. We present in Section~\ref{sec.Fspec} the mechanism for generating hierarchical fermion masses in the model, including the double seesaw mechanism for the neutrinos, and comment on the suppression of lepton flavor violating processes. The mass spectrum of the scalars predicted by the model is presented in detail in Section~\ref{sec.Sc}. Section~\ref{Sec.FCNC} is devoted to an analysis of flavor changing neutral currents. We finish with our conclusions in Section~\ref{conclusion}.

\section{The field content}
\label{model}

Different versions of 3-3-1 models can be defined through the  electric charge operator 
\be\label{Q}
 Q = T_3 + \beta T_8 + X~,
\ee  
where $T_3$ and $T_8$ are the diagonal $SU(3)_L$ generators, and $X$ is the quantum number associated with the Abelian symmetry $U(1)_X$. The parameter $\beta$ characterizes the embedding of the hypercharge operator $Y=\beta T_8 + X$, which commutes with the $SU(2)_L$ generators, within $SU(3)_L\otimes U(1)_X$. In this work, we consider a version with $\beta = -1/\sqrt{3}$~\cite{Singer:1980sw,Montero:1992jk,Foot:1994ym,Hoang:1995vq}, adding right-handed neutrino singlet fields to its minimal fermion content required to cancel the gauge anomalies. Contrary to the cases defined by $\beta = \pm\sqrt{3}$, the current version does not contain fields with exotic electric charges and does not suffer from non-perturbativity issues at low scales~\cite{Dias:2004dc,Dias:2004wk}. 

Taking into account that with respect to SU$(3)_L$ the left-handed fermions are arranged into triplets/antitriplets, while 
the right-handed fermions are singlets, the fermionic multiplets are defined as follows. For the quarks we have two antitriplets -- containing the left-handed quarks corresponding to the first two generations -- and one triplet, plus the corresponding right-handed components in singlets, 
\begin{eqnarray}
&&  Q_{a L} \equiv \begin{pmatrix} d_{aL} \\  -u_{aL} \\ D_{aL} 
\end{pmatrix}_L\sim \left({\bf 3}, {\bf 3^*}, 0 \right),\,\,\,\, Q_{3 L} \equiv \begin{pmatrix} u_{3L} \\  d_{3L} \\ U_{L} 
\end{pmatrix} \sim \left({\bf 3},{\bf 3}, \frac{1}{3} \right), 
 \,\,\,\, \nn\\
&& u^\prime_{n R} = (u_{iR},U_{R})\sim \left({\bf 3},{\bf 1}, 2/3 \right),  \,\,\,\,\,\,  d^\prime_{m R} \ = (d_{iR},D_{aR}) \sim \left({\bf 3},{\bf 1}, -1/3 \right),
\label{qtri}
\end{eqnarray}
where $a = 1,\,2$, $n=1,\,2,\,3,\,4$ and $m=1,\,2,\,3,\,4,\,5$. For the leptons we have three triplets, three right-handed singlets carrying electric charge, plus three right-handed neutral singlets, 
\begin{eqnarray}
& & \Psi_{i L} \equiv \begin{pmatrix} \nu_{iL} \\ e^{-}_{iL}  \\  N_{iL} 
\end{pmatrix}  \sim \left( {\bf 1},{\bf 3}, -\frac{1}{3} \right), \,\,\,\, \nn\\
& & S_{iR}\ \sim \left({\bf 1},{\bf 1}, 0 \right), \,\,\,\,\,  
e^{-}_{iR} \ \sim \left({\bf 1},{\bf 1}, -1 \right)~,
\label{leptri} 
\end{eqnarray}
where $i=1,\,2,\,3$.
The numbers in parentheses above 
describe how these objects transform under the $SU(3)_C$, $SU(3)_L$ and $U(1)_X$ gauge symmetries, respectively. The introduction of the right-handed neutral singlets $S_{iR}$ is not mandatory for cancelling the gauge anomalies, and were not present in the first versions of the model~\cite{Singer:1980sw,Montero:1992jk,Foot:1994ym,Hoang:1995vq}. However, as we will see, such singlets play an important role here in the double seesaw mechanism for generating small masses for the active neutrinos.

The set of scalar fields we consider to break the symmetries contains three triplets of $SU(3)_L$, as in the first versions of the model, 
\begin{eqnarray}\label{sctrip}
 \eta \equiv \begin{pmatrix}
 \eta^{0}_1 \\ \eta^{-}_{2} \\ \eta^0_{3}
 \end{pmatrix} \sim\left(\mathbf{1},\mathbf{3,\,} -\frac{1}{3}\right),\,\,\,\,\,\, 
 \chi \equiv \begin{pmatrix}
 \chi^{0}_1 \\ \chi^{-}_{2} \\ \chi^0_{3}
 \end{pmatrix} \sim\left(\mathbf{1},\mathbf{3,\,} -\frac{1}{3}\right),\,\,\,\,\,\, 
 \rho \equiv \begin{pmatrix}
 \rho^{+}_1 \\ \rho^{0}_{2} \\ \rho^+_{3}
 \end{pmatrix}\sim\left(\mathbf{1},\mathbf{3,\,} \frac{2}{3}\right),
\end{eqnarray}
plus a singlet  
\be \label{sigma}
\sigma \sim ({\bf 1}, {\bf 1},0). 
\ee 
The scalar singlet $\sigma$ plays two important roles in our model. First, it allows the implementation of the solution to the strong CP problem through the Peccei-Quinn mechanism, hosting the axion as a dark matter candidate. Second, through its vacuum expectation value (vev), $\langle \sigma\rangle = v_\sigma/\sqrt{2}$, it takes part in the suppression mechanism leading to hierarchical fermion masses.  

In order to break the $SU(3)_L\otimes U(1)_X$ symmetry down to the electromagnetic factor $U(1)_Q$, along with a consistent mass generation for the fermions, we consider that the components $\chi_3^0$, $\eta_1^0$,  and $\rho_2^0$ acquire different non-vanishing vevs. The vev $\langle \chi_3^0\rangle = w/\sqrt{2}$ leads to the symmetry breakdown $SU(3)_L\otimes U(1)_X$ $\rightarrow$ $SU(2)_L\otimes U(1)_Y$, while the vevs $\langle \eta_1^0 \rangle = v/\sqrt{2}$ and $\langle \rho_2^0 \rangle = u/\sqrt{2}$ perform the symmetry breakdown $SU(2)_L\otimes U(1)_Y$ $\rightarrow$ $U(1)_Q$. Although both the scalar triplets $\eta$ and $\chi$ have each two 
neutral components, we can consider that only $\chi_3^0$ and $\eta_1^0$ get nonzero vevs. One vev of the neutral components in $\chi$ can be eliminated by reparametrization freedom, as discussed in Ref. \cite{Barreto:2017xix},  and the second neutral  component vev in $\eta$ vanishes from the minimization conditions. 

It will be shown that in our scheme $u$ is an effective vev, coming from interactions between the triplet $\rho$ and the other scalar fields in the potential. Thus, we can assume a hierarchy among vevs: $w \gg v \gg u$, with $v \approx 246$ GeV being the electroweak breaking scale. As we will discuss latter, the smallness of $u$ arises from the fact that the field  $\rho$ is taken very massive from the beginning, having a mass $M_\rho\gg w$. In this way this field can be integrated out, and its degrees of freedom will be too heavy to be produced at any ongoing or near-future particle accelerator reaching up to energies of no more than a few TeVs. A direct consequence of this consideration in our construction is that we are left, in principle,  with a reduced number of scalar fields at low energies compared to other 3-3-1 models. 

The $SU(3)_L\otimes U(1)_X$ gauge symmetry gives rise to nine gauge bosons forming physical states  defined through the quadratic terms in  ${\cal L}=\sum_\phi |D_\mu\langle\phi\rangle|^2$, for $\phi=\eta,\,\chi,\,\rho$, with the covariant derivative $D_\mu=\partial_\mu-ig{T_a}W_a-ig_XB_\mu$, where $T_a$, with $a=1,...,8$, are the $SU(3)_L$ generators\footnote{In the fundamental representation $T_a=\frac{\lambda_a}{2}$, where $\lambda_a$ are the Gell-Mann matrices.}, and $g$, $g_X$ the gauge  coupling constants of $SU(3)_L$  and $U(1)_X$, respectively. Four of these gauge bosons correspond to those of the SM: the photon, $Z$, and $W^{\pm}$. The remaining five gauge bosons, denoted by $Z^\prime$, $V^0$, $V^{0\dagger}$, $U^{\pm}$ are supposedly heavier, with their masses related to the scale $w$. 

The non-hermitian gauge bosons can be written as
	\be
	W^{+}_{\mu} = \frac{W_{1\mu}-iW_{2\mu}}{\sqrt{2}},~~ V^{0}_{\mu} =\frac{ W_{4\mu} - iW_{5\mu}}{\sqrt{2}}~~\mbox{and}~~U^{-}_\mu =\frac{W_{6\mu}- iW_{7\mu}}{\sqrt{2}},
	\label{xiv}
	\ee 
and their masses are    
\be
M^2_{W^\pm}\simeq\frac{g^2 v^2}{4},~~~  M^{2}_{V^0}=M^{2}_{({V^0})^\dag}= \frac{g^2}{4}(v^2+w^{2})~~\mbox{and}~~M^2_{U^\pm}\simeq\frac{g^2 w^2}{4}~, 
\ee
where contributions coming from $u \ll v, w$ have been neglected. This implies a peculiar tree level relation $M^{2}_{V^0}-M^2_{U^\pm}\simeq M^2_{W^\pm}$. 
We can see that this mass splitting coincides with the one in Ref. \cite{Barreto:2017xix}, but the ordering is opposite: here $V^0$ is heavier. This fact could be used to distinguish both models if, eventually, the new gauge bosons were discovered.

The massless field $A_\mu$, associated with the photon, is given by
\begin{equation} \label{photon}
 A_{\mu} = \frac{\sqrt{3}}{\sqrt{3+4t^2}}\left( t \  W^{3}_{\mu}-\frac{t}{\sqrt{3}} \ W^{8}_{\mu}+  B_{\mu}\right)\nn~.
\end{equation}
Finally, the last two physical fields in the gauge sector spectrum are the massive neutral gauge bosons $Z^1_\mu$ and $Z^2_\mu$. These fields can be conveniently written in terms of $Z_\mu$ and $Z^\prime_\mu$, where the former is the SM $Z$ boson, and the latter is associated with the 3-3-1 symmetry breaking down to the SM group,
\begin{equation}
\left(
\begin{array}{c}
Z^1   \\ 
Z^2 
\end{array}
\right)=
\left(
\begin{array}{cc}
{\rm cos}\,\varphi & -{\rm sin}\,\varphi   \\ 
{\rm sin}\,\varphi  & {\rm cos}\,\varphi  
\end{array}
\right)
\left(
\begin{array}{c}
Z   \\ 
Z^\prime 
\end{array}
\right).
\label{zzl}
\end{equation}
The mixing matrix above diagonalizes the following mass matrix, written in the basis $(Z_\mu,~Z^\prime_\mu)$, 
\begin{equation}
\mathcal{M}_Z = 
\left(
\begin{array}{cc}
M_Z^2  & M_{Z\,Z^\prime}^2   \\ 
M_{Z\,Z^\prime}^2  & M_{Z^\prime}^2  
\end{array} 
\right)\,,
\label{zzlmm}
\end{equation}
where
\begin{eqnarray}
& & M_Z^2=\frac{g^2 v^2}{4 \, \cos^2(\theta_W)}\,, \hspace{1 cm} 
M_{Z\,Z^\prime}^2 =-M_Z^2 \frac{\cos(2 \theta_W)}{\sqrt{3-4 \, \sin^2(\theta_W)}}\,,
\nonumber\\
& & M_{Z^\prime}^2 = \frac{M_Z^2 \cos(2 \,\theta_W) + g^2w^2\cos^2(\theta_W)}{{3-4 \, \sin^2(\theta_W)}}\,,
\end{eqnarray}
with $\sin^2(\theta_W) \approx 0.231$. 
Thus, the mixing angle $\varphi$ between the SM $Z$ boson and $Z^\prime$ is 
\begin{equation}\label{ZZmix}
\tan (2\,\varphi)=\frac{2M_{Z\,Z^\prime}^2}{M_{Z^\prime}^2-M_Z^2}.
\end{equation}
For $w=10$ TeV ($M_{Z^\prime} \approx 4$  TeV), for example, we obtain $\varphi \approx -  10^{-4}$ in such a way that for most analyses it is enough to take $Z^1_{\mu}= Z_\mu$ and $Z^2_{\mu}= Z^\prime_\mu$. %The definition of these states will be important for the discussion on flavor changing neutral currents; see Sec.\,\ref{Sec.FCNC}.

\section{Fermion masses}\label{sec.Fspec}

In this section, we explore the generation of mass to all fermions in the model. We divide it in three main steps. First, we consider the renormalizable interactions of the fermion fields with two scalar triplets only: $\eta$ and $\chi$. Second, after showing that this configuration is not enough to render all fermions massive, we include effective operators coming from the integration of the heavy scalar triplet $\rho$ and show that all fermions become massive. At this point, although no massless fermion remains, the mass hierarchies provided by the model do not reproduce naturally the experimental results. This issue is then dealt with in the third step with the imposition of a discrete symmetry along with the introduction of the scalar singlet $\sigma$.

We start by writing down all renormalizable Yukawa terms allowed by the gauge symmetries involving all fermions and two scalar triplets, $\eta$ and $\chi$,
\begin{eqnarray}
-\label{Yuk}
\mathcal{L}_Y &=& h^{\nu}_{ij}\overline{\Psi_{iL}}~\eta ~S_{jR}+h^{N}_{ij}\overline{\Psi_{iL}}~ \chi~S_{jR} +\frac{1}{2}\mu_{ij} \overline{S_{iR}^c} S_{jR} \nn\\  
&+& h^{d}_{am}\overline{Q_{a L}}~ \eta^*d_{m R}^\prime+h^{D}_{am}\overline{Q_{a L}}~ \chi^*d_{m R}^\prime \nn\\  
&+& h^{u}_{n}\overline{Q_{3 L}}~ \eta\, u_{n R}^\prime + 
h^{U}_{n}\overline{Q_{3 L}}~ \chi\, u_{n R}^\prime+ h.c.~,
\end{eqnarray}
where $h^\nu_{ij}$, $h^N_{ij}$ and $\mu_{ij}$, or simply ${\bf h}^\nu, {\bf h}^N$, $\bs{\mu}$, are $3\times 3$ complex matrices;  $h^d_{am}, h^D_{am}$ (${\bf h}^d, {\bf h}^D$) are $2\times 5$ matrices; and $h^u_{n}, h^U_{n}$ (${\bf h}^u, {\bf h}^U$) are $1\times 4$ matrices.
As shown previously in different 3-3-1 versions, when considering a minimal scalar sector containing only two triplets, some fermions remain massless due to the presence of a residual Peccei-Quinn-like (PQL) symmetry \cite{Barreto:2017xix, Montero:2014uya}. In the current case, considering the operators in Eq. (\ref{Yuk}) with the first component of $\eta$ and the third component of $\chi$ acquiring non-vaninshing vevs, one can see that the charged leptons, the up-type quarks of the first two families, and the down-type quark of the third family  do not get mass terms.

In order to generate masses to all fermions, the global PQL symmetry must be broken explicitly. This step can be achieved with the introduction of the following non-renormalizable dimension-5 operators, suppressed by an energy scale $\Lambda\gg w$, 
\bea
-\mathcal{L}_5 &= & \frac{y^\nu_{ij}}{\Lambda} [\overline{\Psi_{iL}}~\Psi_{jL}^c][ \chi\,\eta]+\frac{y^e_{ij}}{\Lambda}
\overline{\Psi_{i L}} [\chi\,\eta]^* e_{j R}  \nn \\
&+&\frac{y^d_{m}}{\Lambda} \overline{Q_{3 L}}\, [\chi\,\eta]^* d_{m R}^\prime
+\frac{y^u_{an}}{\Lambda} \overline{Q_{a L}}\, [\chi \,\eta]u_{n R}^\prime +h.c.,
\label{D5}
\eea 
where the terms between brackets should be understood as the antisymmetric product of the respective $SU(3)_L$ triplets, whose components are, for example,  $[\chi\,\eta]_p\equiv\epsilon_{pqr}\chi_q\eta_r$,  with $p,\,q,\,r=1,\,2,\,3$. The coupling matrices in the Lagrangian above can be classified as $3\times 3$ matrices: $y^\nu_{ij}, y^e_{ij}$ (${\bf y}^\nu, {\bf y}^e$); a $2\times 4$ matrix:  $y^u_{an}$ (${\bf y}^u$); and a $1\times 5 $ matrix: $y^d_m$ (${\bf y}^d$). 
Additionally, $\mathbf{y}^\nu$ is antisymmetric.\,\footnote{%
As an effective operator, a symmetric piece could be present but it will not arise from integrating out $\rho$; see Sec.\,\ref{rhoint}.
}
Similarly to the mechanism proposed in Ref.~\cite{Barreto:2017xix}, in Sec. \ref{sec.Sc} we show that the effective operators in Eq. (\ref{D5}) can eventually emerge considering that, differently from the triplets $\eta$ and $\chi$, the  scalar triplet $\rho$ defined in Eq. (\ref{sctrip}) has a mass term ${\cal{L}}\supset -M^2_\rho\rho^\dagger\rho$, with $M_\rho\gg w$. Thus, at lower energies ($\sim w$) we have that $\rho\simeq [\chi\,\eta]/\Lambda$ so that this field can be integrated out leading to the effective operators in Eq. (\ref{D5}). The energy scale $\Lambda$ is related to the mass $M_\rho$ and the vev $v_\sigma$ of the scalar singlet as we will see. A small effective vev is then generated for the neutral component of $\rho$,  
\be\label{effvev} 
\langle \rho_2^0 \rangle = \frac{v w}{2\Lambda} =  \frac{u}{\sqrt{2}}.
\ee

When the Lagrangians in Eqs. (\ref{Yuk}) and (\ref{D5}) are taken into account, and the scalar triplets acquire non-vanishing vevs, the mass matrices below are generated making all fermions massive.
\begin{itemize}

\item $3\times3$ charged lepton mass matrix:  
\be\label{clmm}
\mathcal{M}^e = \frac{u  }{\sqrt{2}}~ {\bf y}^{e}.
\ee 
If we identify the energy scale $u$ with the mass of the heaviest charged lepton, the tau, we have $u \sim m_\tau \sim 1$ GeV, implying that $\Lambda \sim 10^6$ GeV with $v\sim 10^2$ GeV and $w\sim 10^4$ GeV. To obtain the correct masses for the lighter charged leptons, the muon and the electron, suppressed couplings in ${\bf y}^e$ are required. When comparing to the SM case, where the charged fermion masses are proportional to $v_{EW}$, instead of $u$, our model requires less suppression of the Yukawa constants.

\item $4\times4$ up-type quark mass matrix written in the basis ($u_a, u_3, U$): 
\be \label{uqmm1}
\mathcal{\tilde{M}}^u =\frac{1}{\sqrt{2}}\begin{pmatrix}
- u {\bf y}^u   \\  v  {\bf h}^{u} \\ w  {\bf h}^{U}  
\end{pmatrix}~.
\ee
For the up-type quarks, we see that the first two SM families get masses proportional to $u$, the mass of the top quark is proportional to $v$, and the new quark mass is proportional to $w$. Thus, the present model provides a more natural way of explaining the mass hierarchy between the third and the other two families than the SM.

\item $5\times5$ down-type quark mass matrix in the basis ($d_a, d_3, D_a$):
\be\label{dqmm1}
\mathcal{\tilde{M}}^d =\frac{1}{\sqrt{2}}\begin{pmatrix}
v  {\bf h}^{d}\\ u {\bf y}^d \\ w {\bf h}^{D} 
\end{pmatrix}~.
\ee
In this case, however, we notice an inverted hierarchy, since the first two down-quark families have masses proportional to $v$, while the third, which should be heavier, gets a mass proportional to $u$. 

\end{itemize}

Having two neutral fermion fields in each lepton triplet $\Psi_{iL}$ plus three fermionic neutral singlets $S_{iR}$, the model can feature a double seesaw mechanism for neutrino mass generation~\cite{Mohapatra:1986aw,Mohapatra:1986bd,Barr:2003nn}.

\begin{itemize}
\item The neutrino mass matrix, in the flavor basis $(\nu_{iL}, N_{iL}, S_{iR}^c)$ with convention $\overline{\psi_L}\psi_L^c$, is given by
\be\label{nemm1}
\mathcal{\tilde{M}}^\nu = \begin{pmatrix}
 0 & {\bf m_{L}} &\mathbf{m_D^\nu} \\
 ({\bf m_L})^T & 0 &\mathbf{m_D^{\mathit{N}}} \\
 (\mathbf{m_D^\nu})^T & (\mathbf{m_D^{\mathit{N}}})^T & \bs{\mu}
\end{pmatrix}~,
\ee 
with $\sqrt{2} {\bf m_{L}} =-2u\,{\bf y}^{\nu}$,
$\sqrt{2} \mathbf{m_D^\nu} = v\,{\bf h}^{\nu}$, $\sqrt{2} \mathbf{m_D^{\mathit{N}}}= w\,{\bf h}^{N}$ 
and $\bs{\mu}$ was defined in \eqref{Yuk},
all of which are $3 \times 3$ matrices. With this texture, the double seesaw takes place naturally when $\mu\gg w\gg v,u$, where $\mu$ is the order of magnitude of $\bs{\mu}$.
The lightest neutrinos, i.e., the active ones, will get the following dominant contribution to its mass matrix
\bea 
{\bf M}_{\nu} &\simeq& {\bf m_L}\, [(\mathbf{m_D^\mathit{N}})^T]^{-1} \,\bs{\mu}\, (\mathbf{m_D^{\mathit{N}}})^{-1} \,{\bf m_L}^T \nn\\ 
&\sim& 2\,{\bf y}^{\nu}\, [({\bf h}^{N})^T]^{-1} \,\bs{\mu}({\bf h}^{N})^{-1}\, ({\bf y}^{\nu})^T\times 10^{-8}~.
\eea
Therefore, to get down to the sub-eV scale for the active neutrino masses, with $\mu\gg w\sim 10^4$ GeV, a large amount of suppression of the Yukawa couplings will be required. When assuming that $\mu = 10^8$ GeV, for example, one way of getting light enough neutrinos is to take the coefficients in 
${\bf y}^\nu$ to be no larger than $10^{-4}$, for ${\bf h}^N$ of order one. 
Note that one neutrino is automatically massless within this approximation because of the antisymmetry of $\mathbf{y}^\nu$.

\end{itemize}

In the next subsections, we implement the third step: the imposition of a discrete symmetry and the introduction of the scalar singlet $\sigma$  that breaks it down spontaneously. With these new ingredients, in Sec. \ref{ssec.Hier}, we show how to obtain hierarchical masses to all charged fermions and, in Sec. \ref{ssec.doub}, 
how suppressed Yukawa couplings naturally arise to generate the correct mass scale for active neutrinos.
As added bonuses, we observe that the model now counts with an invisible axion which solves the strong CP problem and plays the role of cold dark matter.

\subsection{Hierarchical quark masses}\label{ssec.Hier}

We start this section by assuming that in addition to the gauge symmetries, our model is also invariant under a discrete $\ZZ_9$ symmetry. Under this discrete symmetry, the fermion and scalar fields, including the scalar singlet $\sigma$ defined in Eq. (\ref{sigma}), transform as described in Table \ref{TZ}. 
\begin{center}
	\begin{table}[h]
		\begin{tabular}{ |c|c|c|c|c|c|c|c|c|c|c|c|c|c| } 
			\hline
            {\bf Fields}  &  $\eta$  & $\chi$  & $Q_{aL}$ &  $Q_{3L}$ & $u_{aR}$ & $u_{3R}$  & $U_R$ & $d_{jR}$ & $D_{aR}$ & $e_{iR}$ &  $S_{iR}$ &    $\Psi_{iL}$ & $\sigma $ \\\hline 
			  $ \ZZ_9 $ & $ \omega^{-4}$ & $ \omega^{2} $ & $ \omega^2 $ & $ \omega^{-2}  $  & $ \omega^{4} $ & $ \omega^2  $ &  $ \omega^{-4}  $  & $ \omega^4  $ & $ \omega^{-4} $ & $ \omega^{-1} $ & $ \omega^{-1} $ & $ \omega^{1}  $ & $\omega^2$  \\  \hline
		\end{tabular}
		\caption{$\ZZ_{n}$ transformations with $\omega = \exp\left(\frac{2 \pi i}{n}\right)$.}
        \label{TZ}
	\end{table}
\end{center}

Such a discrete symmetry forbids some of the previous Yukawa interactions, for example, the term $h^d_{ab} \overline{Q_{aL}} \eta^* d_{bR}$ in Eq. (\ref{Yuk}) which attributed to the first two families of down-type quarks a mass term proportional to $v$, leading to an inverted hierarchy with the third family. The terms that survive in Eqs. (\ref{Yuk}) and (\ref{D5}) are
\begin{eqnarray}\label{YukZ}
-\mathcal{L} &=& h^{N}_{ij}\,\overline{\Psi_{iL}}\,\chi\,S_{jR} + h^{D}_{a,3+b}\,\overline{Q_{a L}}\, \chi^* \,D_{b R}+ 
h^{u}_3\,\overline{Q_{3 L}}\, \eta\, u_{3 R}+ 
h^{U}_4\,\overline{Q_{3 L}}\, \chi\, U_{R}\nn\\  
&&+\frac{y^e_{ij}}{\Lambda}\,
\overline{\Psi_{i L}} \,[\chi\,\eta]^*\, e_{j R}
+\frac{y^d_j}{\Lambda} \,\overline{Q_{3 L}}\,[\chi\,\eta]^*\,  d_{j R}
+\frac{y^u_{ab}}{\Lambda} \,\overline{Q_{a L}}\,[\chi\,\eta]\,  u_{b R}  +h.c.~,
\end{eqnarray}
where $h^N_{ij}$ and $y^e_{ij}$ are the same as before: $3 \times 3$ matrices; $h^D_{a,3+b}$ and $y^u_{ab}$ are $2\times2$ matrices; $y^d_{j}$ is a $1\times3$ matrix; $h^u_3$ and $h^U_4$ are complex numbers. 

The operators forbidden with the imposition of the discrete symmetry can now reappear multiplied by the appropriate power of $\sigma$ (or $\sigma^*$):
\begin{eqnarray}\label{Yuksig}
-\mathcal{L}^{(\sigma)} &=& \frac{1}{2}\tilde{h}^S_{ij}\, \sigma\, \overline{S_{iR}^c} \,S_{jR}+{\tilde h}^{\nu}_{ij}\,\left(\frac{\sigma}{\Lambda^\prime}\right)^3\, \overline{\Psi_{iL}}\,\eta \,S_{jR}+\frac{{\tilde y}^\nu_{ij}}{\Lambda }\,\left(\frac{\sigma}{\Lambda^\prime}\right)^2\,[ \overline{\Psi_{iL}}\,\Psi_{jL}^c]\,[\chi\,\eta]\,\nn\\
&&+{\tilde h}^{d}_{aj}\,\left(\frac{\sigma}{\Lambda^\prime}\right)\,\overline{Q_{a L}}\, \eta^*d_{j R}+{\tilde h}^{d}_{a,3+b}\,\left(\frac{\sigma^*}{\Lambda^\prime}\right)^3\,\overline{Q_{a L}}\, \eta^*D_{b R} +{\tilde h}^{D}_{aj}\,\left(\frac{\sigma}{\Lambda^\prime}\right)^4\,\overline{Q_{a L}}\,\chi^*d_{j R}\nn\\
&&+ \frac{{\tilde y}^{d}_{3+a}}{\Lambda}\,\left(\frac{\sigma^*}{\Lambda^\prime}\right)^4\,\overline{Q_{3 L}}\,[\chi\,\eta]^*\, D_{a R}+{\tilde h}^{u}_{a}\,\left(\frac{\sigma^*}{\Lambda^\prime}\right)\, 
\overline{Q_{3 L}}\, \eta\, u_{a R} + {\tilde h}^{u}_4\,\left(\frac{\sigma}{\Lambda^\prime}\right)^3 \,\overline{Q_{3 L}}\, \eta\, U_{R}\nn\\
&&+ {\tilde h}^{U}_a\,\left(\frac{\sigma^*}{\Lambda^\prime}\right)^4\,\overline{Q_{3 L}}\, \chi\, u_{a R}+ {\tilde h}^{U}_3\,\left(\frac{\sigma^*}{\Lambda^\prime}\right)^3\,\overline{Q_{3 L}}\, \chi\, u_{3 R} + \frac{{\tilde y}^u_{a3}}{\Lambda}\,\left(\frac{\sigma }{ \Lambda^\prime}\right) \,\overline{Q_{a L}}\,[\chi\,\eta]^*\, u_{3 R}\nn\\
&&+ \frac{{\tilde y}^u_{a4}}{\Lambda}\,\left(\frac{\sigma }{ \Lambda^\prime}\right)^4\, \overline{Q_{a L}}\,[\chi\,\eta]^*\, U_{R} +h.c.~,
\end{eqnarray}
where $\Lambda'$ is a large mass scale, the largest appearing in our model, suppressing the higher-dimensional operators, and the Yukawa couplings assume the following forms: ${\tilde h}^{S}_{ij}$, ${\tilde h}^{\nu}_{ij}$ and ${\tilde y}^\nu_{ij}$ ($\mathbf{\tilde{h}^S}$, $\bf{{\tilde h}}^{\nu}$ and $\bf{{\tilde y}}^\nu$) are $3\times3$ matrices; ${\tilde h}^{d}_{aj}$ and ${\tilde h}^{D}_{aj}$ are $2\times3$ matrices; ${\tilde h}^{d}_{a,3+b}$ is a $2\times2$ matrix; ${\tilde y}^{d}_{3+a}$, ${\tilde h}^{u}_a$ and ${\tilde h}^{U}_a$ are $1\times2$ matrices; ${\tilde y}^u_{a3}$ and ${\tilde y}^u_{a4}$  are $2\times1$ matrices; ${\tilde h}^u_4$ and ${\tilde h}^U_3$ are complex numbers. Except for the first operator, all the others have mass dimension superior to 4 and, consequently, $\sigma$ (or $\sigma^*$) appears suppressed by $\Lambda'$, with $v_\sigma/\Lambda'\ll 1$.

Upon analyzing Eqs. (\ref{YukZ}) and (\ref{Yuksig}), it is possible to see that three Abelian symmetries are present. Two of them are the gauged $U(1)_X$ and the global $U(1)_B$ associated with the Baryon number. The other one is a Peccei-Quinn symmetry $U(1)_{PQ}$ under which the fields have the charges shown in Table \ref{TPQ}.
\begin{center}
	\begin{table}[h]
		\begin{tabular}{ |c|c|c|c|c|c|c|c|c|c|c|c|c|c| } 
			\hline
            {\bf Fields}  & $\sigma$ &  $\eta$  & $\chi$  & $Q_{aL}$ &  $Q_{3L}$ & $u_{aR}$ & $u_{3R}$  & $U_R$ & $d_{jR}$ & $D_{aR}$ & $e_{iR}$ &  $S_{iR}$ &    $\Psi_{iL}$ \\ \hline 
			  $ X_{PQ} $ & $ 1 $ & $ 0 $ & $ 3 $ & $ 0  $  & $ -4 $ & $ -3  $ &  $ -4  $  & $ -7  $ & $ -1 $ & $ 3 $ & $ 11/2 $ & $ -1/2  $ & $ 5/2 $ \\  \hline
		\end{tabular}
		\caption{$U(1)_{PQ}$ charges}
        \label{TPQ}
	\end{table}
\end{center}

The solution to the strong CP problem in our model is provided by the anomalous feature of the $U(1)_{PQ}$ symmetry above, given by the non-vanishing of  the color anomaly coefficient. Taking into account the PQ charges in Table \ref{TPQ}, such coefficient is $C_{ag} =\sum_{i=\text{quarks}} (X_{iL}-X_{iR})= 2$, and also enters in the definition of the axion decay constant\footnote{The axion decay constant is defined through the normalization of the axion kinetic term, and in the present model it is $f_a=\sqrt{X_\eta^2 v^2+X_\chi^2 w^2+X_\sigma^2 v_\sigma^2 }/|C_{ag}|\approx v_\sigma/|C_{ag}|$, since $v_\sigma \gg w \gg v$. The model has a domain wall number equal to $N_{DW}= |C_{ag}|=2$. One can see this by computing $C_{ag}$  in the normalization where all the $U(1)_{PQ}$ charges are integers and observing that although in this case the axion potential is invariant under a discrete $\ZZ_4\subset U(1)_{PQ}$ symmetry there is a discrete $\ZZ_2$ subgroup acting trivially on the vacuum. For more details see, for example, the appendix of Ref.~\cite{Dias:2014osa}.} $f_a\approx v_\sigma/|C_{ag}|$. The $U(1)_{PQ}$ symmetry is spontaneously broken when the scalar singlet gets a vev $\langle\sigma\rangle=v_\sigma/\sqrt{2}$. This, in turn, will give rise to the axion field, the pseudo Nambu-Goldstone boson of the $U(1)_{PQ}$ symmetry, which gets mass via non-perturbative effects.  When considering $v_\sigma \gg w \gg v$, the axion will be invisible -- due the suppression of its couplings by $1/v_\sigma$ -- and mostly composed of the imaginary part of $\sigma$, as in the original invisible axion models~\cite{Kim:1979if,Shifman:1979if,Dine:1981rt,Zhitnitsky:1980tq}. For the singlet vev in the interval $10^9\text{ GeV} \lesssim f_a\lesssim 10^{13} \text{ GeV}$ the axion could also play the role of dark matter~\cite{Tanabashi:2018oca}. 

Let us now discuss the fermion masses in our model by turning our attention back to Eqs.~(\ref{YukZ}) and (\ref{Yuksig}). The operator behind the charged lepton masses has not been altered by the discrete symmetry, therefore, their masses are still given by Eq. (\ref{clmm}), and the discussion below such an equation remains valid. The new ingredients have consequences to the quark masses, and their new mass matrices are shown below. In summary, we now have
\begin{itemize}

\item a $5\times5$ down-type quark mass matrix:
\be\label{dqmm2}
\mathcal{M}^d =\frac{1}{\sqrt{2}}\begin{pmatrix}
\kappa v  \,{\tilde h}_{11}^{d} &\kappa v\,  {\tilde h}_{12}^{d} & \kappa v \, {\tilde h}_{13}^{d} & \kappa^3 v \, {\tilde h}^{d}_{14} & \kappa^3 v \, {\tilde h}^{d}_{15} \\ 
\kappa v \, {\tilde h}_{21}^{d} &\kappa v \, {\tilde h}_{22}^{d} & \kappa v \, {\tilde h}_{23}^{d} & \kappa^3 v \, {\tilde h}^{d}_{24} & \kappa^3 v \, {\tilde h}^{d}_{25} \\ u\,  y_1^d & u  \,y_2^d &u \, y_3^d & \kappa^4 u \, {\tilde y}_4^d & \kappa^4 u \, {\tilde y}_5^d \\ \kappa^4 w \,{\tilde h}_{11}^{D} &  \kappa^4w\, {\tilde h}_{12}^{D} &  \kappa^4 w \,{\tilde h}_{13}^{D} & w \, h_{14}^{D} & w \, h_{15}^{D} \\ \kappa^4 w \, {\tilde h}_{21}^{D} &  \kappa^4 w \, {\tilde h}_{22}^{D} &  \kappa^4 w \, {\tilde h}_{23}^{D} & w \, h_{24}^{D} & w \, h_{25}^{D} 
\end{pmatrix}~,~~~ \mbox{with}~~~~\kappa = \frac{v_\sigma}{\sqrt{2}\Lambda^\prime}.
\ee
We see now that the hierarchy has changed. Instead of being proportional to $v$, the $2\times2$ upper-block appears multiplied by the suppression factor $\kappa$. We have the freedom to choose $\kappa v\sim m_s \sim 10^{-1}$ GeV, where $m_s$ is the mass of the strange quark, in such a way that $\kappa \sim 10^{-3}$. Furthermore, as before, the bottom quark mass is already proportional to the natural scale, {\it i.e.}, $m_b\sim u\sim 1$ GeV. The new quarks become heavy with masses at the $w$ scale, and their mixing with the standard down-type quarks are suppressed by powers of $\kappa$. Thus, our model now accounts for the correct mass hierarchy among the SM down-type quarks, and, in addition, it effectively decouples the standard quarks from the exotic ones.

\item a $4\times4$ up-type quark mass matrix: 
\be \label{uqmm2}
\mathcal{M}^u =\frac{1}{\sqrt{2}}\begin{pmatrix}
- u \, y^u_{11} & - u\, y^u_{12}  & - \kappa u\,  {\tilde y_{13}}^u & - \kappa^4 u \,\tilde{ y}_{14}^u \\- u \, y^u_{21} & - u \, y^u_{22}  & - \kappa u\,  {\tilde y_{23}}^u & - \kappa^4 u\, \tilde{y}_{24}^u \\ \kappa v \, {\tilde  h}^{u}_{1} & \kappa v \, {\tilde  h}^{u}_{2} & v \, h^{u}_3 & \kappa^3 v  \, \tilde{h}^{u}_4 \\ \kappa^4 w \, \tilde{h}_{1}^{U} & \kappa^4 w \, \tilde{h}_{2}^{U} & \kappa^3 w \, {\tilde h}^{U}_3 & w \, h^{U}_4  \\
\end{pmatrix}~.
\ee
The previous mass hierarchy presented in Eq. (\ref{uqmm1}) is preserved: the first two SM families get masses proportional to $u$, the mass of the top quark is proportional to $v$, and the new quark mass is proportional to $w$. The main difference lies on the fact that the mixing between the standard and the new up-type quarks becomes negligible due to the suppression by many powers of $\kappa$. Once again, our model provides a more natural hierarchy among the up-type quark masses than the SM case.

\end{itemize}

Keeping track only of the orders of magnitude, we can represent the structure of the mass matrices for the quarks in Eqs. (\ref{dqmm2}) and (\ref{uqmm2}) as
\eq{
\label{quark:mass:pattern}
\mathcal{M}^d/u\sim 
\left(
\begin{array}{c|c|c}
0.1 & 0.1 & 10^{-7}\cr
\hline
1 & 1 & 10^{-12}\cr
\hline
10^{-8} & 10^{-8} & 10^{4}\cr
\end{array}
\right),\quad
\mathcal{M}^u/u\sim 
\left(
\begin{array}{c|c|c}
1 & 10^{-3} & 10^{-12}\cr
\hline
0.1 & 10^2 & 10^{-7}\cr
\hline
10^{-8} & 10^{-8} & 10^{4}\cr
\end{array}
\right),
}
where the upper-left block always refers to the first two families and the central block to the third family.
This structure clearly shows that the mixing amongst the
standard and new quarks is very suppressed by powers of $\kappa \sim 10^{-3}$. Such matrices, therefore, are effectively block-diagonal, and flavor changing effects related to this mixing are expected to be negligible; see Sec.\,\ref{Sec.FCNC}.

\subsection{Neutrino masses: double seesaw mechanism}\label{ssec.doub}

Let us now discuss the mechanism behind the neutrino mass generation in our model. The texture of the mass matrix in Eq. (\ref{nemm1}) remains valid but
now some of the $3\times 3$ matrices are modified as
\eq{\label{Mnumass}
\sqrt{2} {\bf m_{L}}= -2\kappa^2 u {\bf {\tilde y}}^{\nu}, 
\quad
\sqrt{2} \mathbf{m_D^\nu} = \kappa^3 v  {\bf {\tilde h}}^{\nu}, 
\quad
\sqrt{2}\bs{\mu} = v_\sigma {\bf {\tilde h}}^{S}
\,.
} 
The matrix $\mathbf{m_D^{\mathit{N}}}$ remains the same.
We can see that the first two matrices have their magnitudes greatly suppressed by powers of $\kappa$ for order one couplings, and the scale $\mu$ is linked to the PQ breaking scale.

As $\bs{\mu} \gg \mathbf{m_D^{\mathit{N}}} \gg  {\bf m_L}, \mathbf{m_D^\nu}$, a double seesaw takes place. The physical spectrum is comprised of
three very heavy neutrinos whose main contribution comes from $S_{iR}$ and with masses now proportional to the PQ breaking scale,
\be \label{Smass}
\mathbf{M_S} \simeq  \bs{\mu}
~,
\ee
three intermediate scale neutrinos whose main contribution comes from $N_{iL}$ and with masses
\be\label{Nmass} 
\mathbf{M_N} \simeq -\mathbf{m_D^{\mathit{N}}}\, \bs{\mu}^{-1} (\mathbf{m_D^{\mathit{N}}})^T
,
\ee
and, finally, the active neutrinos $\nu_{iL}$ are required to have sub-eV masses naturally,
\be\label{numass} 
\mathbf{M_{\nu}} \simeq
- \mathbf{m_L}\mathbf{M_N}^{-1} \mathbf{m_L}^T 
\sim 0.1~\unit{eV}~.
\ee 

Considering that $\mu\lesssim v_\sigma\sim 10^{10}\,\unit{GeV}$, $\mathbf{m_D^{\mathit{N}}}\lesssim w\sim 10^4\,\unit{GeV}$, $\mathbf{m_L}\lesssim \kappa^2u\sim 10^{-6}\,\unit{GeV}$, $\mathbf{m_D^{\nu}}\lesssim \kappa^3 v\sim 10^{-7}\,\unit{GeV}$,
we can choose as representative scales
\eq{
\label{M:SN}
\mathbf{M_{S}}\sim 10^8\,\unit{GeV},
\quad
\mathbf{M_{N}}\sim \unit{GeV}\,,
}
which is valid for $\mathbf{m_D^{\mathit{N}}}\sim w$ and implies $\mathbf{m_L}\sim 10^{-5}\,\unit{GeV}$, which is still natural by choosing $\kappa$ slightly larger than $10^{-3}$ and $\tilde{y}^\nu$ larger than unity.
In contrast, we have to choose $\mu$ to be somewhat lower than $v_\sigma = 10^{10}$ GeV by suppressing the coupling $\tilde{h}^S \sim 10^{-2}$ to increase the scale of the intermediate neutrinos close to the GeV scale.
Much lighter intermediate neutrinos that mix with $\nu_e$ may lead to problems during BBN and direct detection constraints\,\cite{deppisch}. We briefly detail these aspects below.
The choice of PQ scale is dictated so that the axion solves the strong CP problem and is also a dark matter candidate.

Let us discuss the constraints on GeV and sub-GeV intermediate neutrinos.
Many effects depend on the mixing between these neutrinos and active neutrinos which can be quantified as\,\cite{Hettmansperger:2011bt}
\be \label{nuNmixing}
U_{\nu N} \simeq 
m_L M_N^{-1}U_{PMNS}
\sim 10^{-5}
~,
\ee 
for $U_{PMNS}\sim 1$. The last number follows from our choice in Eq. \eqref{M:SN}.
For sterile neutrinos that mix with $\nu_e$ the strongest constraints come from the limit of neutrinoless double beta decay which prefers lower mixing and BBN constraints in standard cosmological scenarios\,\cite{Ruchayskiy:2012si, Vincent:2014rja, Adhikari:2016bei}
which prefer larger mixing.
For example, for a 0.7 GeV neutrino, the mixing needs to be restricted to $10^{-5}~\text{--}~10^{-4}$\,\cite{deppisch}.
For larger masses, the interval widens and our choice is phenomenologically viable. If the mixing with the $e$ flavor is further suppressed, these constraints become much weaker.

Lepton flavor violating processes are also very suppressed.
For example, the branching ratio for the flavor changing decay $\mu\to e\gamma$ induced by $N_i$ exchange is given by\,\cite{mu->e.gamma}
\be
\mathrm{Br}(\mu\to e\gamma)=
\frac{\alpha_w^3s_w^2}{256\pi^2}\frac{m_\mu^4}{m_W^4}\frac{m_\mu}{\Gamma_\mu}
\left|\sum_i
U_{eN_i}U^*_{\mu N_i}G_{\gamma}(x_{N_i})
\right|^2\,,
\ee
where $G_\gamma$ is a loop function and $x_{N_i}=M^2_{N_i}/m_W^2$.
The pre-factor contributes $4\times 10^{-3}$ while the loop function $G_\gamma\sim x_{N_i}/4\sim 4\times 10^{-5}$ for our GeV $N_i$,
and then, taking into account the tiny mixing between these intermediate scale neutrinos and the active ones given in Eq. (\ref{nuNmixing}), the branching ratio is far below the current bound of $6\times 10^{-13}$.
Another contribution from a similar diagram with $U^+$ and $N$ in the loops is even more suppressed.

At last, the small tuning to get a small scale $\mu$ may be not necessary in non-standard cosmological scenarios, such as the low reheating scenarios of Ref.~\cite{Gelmini:2008fq}.
In this case we can have sub-GeV or MeV neutrinos in our model with $\mu\sim v_\sigma$.

\section{The scalar sector}\label{sec.Sc}

We consider here the scalar potential made out of all the scalar fields: three $SU(3)_L$ triplets $\eta$, $\chi$ and $\rho$, as defined in Eq. (\ref{sctrip}), plus the scalar singlet $\sigma$ from Eq. (\ref{sigma}). Some of the scalar fields will become massive, while others will be absorbed by the gauge sector via the Higgs mechanism. These features are studied in this section, and the scalar spectrum is presented.

In order to find the scalar spectrum, we decompose the complex neutral fields that acquire a non-vanishing vev into their scalar, $S_\varphi$, and pseudoscalar, $A_\varphi$, components
\bea\label{sctripdec}
\chi = \begin{pmatrix} \chi_1^0\\ \chi_2^{-} \\ 
\frac{1}{\sqrt{2}}(w + S_{\chi} + i A_{\chi} )
\end{pmatrix}
\,,~~
\eta = \begin{pmatrix}
\frac{1}{\sqrt{2}}( v+ S_{\eta} + i A_{\eta} )\\ \eta_2^{-} \\ 
\eta_3^0
\end{pmatrix}
\,,~~
\rho = \begin{pmatrix}
\rho_1^+\\ \frac{1}{\sqrt{2}}( u+ S_{\rho} + i A_{\rho} ) \\ 
\rho_3^+
\end{pmatrix}\,,
\eea	
and
\be\label{sigmadec}
\sigma = \frac{1}{\sqrt{2}}( v_\sigma + S_{\sigma} + i A_{\sigma} ).
\ee 
The hierarchy of vevs obeys $v_\sigma\gg w\gg v\gg u$.

Taking into account that the heavy triplet $\rho$ transforms trivially under the imposed discrete symmetry, while the other fields transform according to Table \ref{TZ}, we can write down the most general renormalizable scalar potential as
\bea \label{V}
V &=& \mu_\eta^2 \,|\eta|^2 + M_\rho^2 \,|\rho|^2 + \mu_\chi^2 \,|\chi|^2 + \mu_\sigma^2 \,|\sigma|^2 + \lambda_\eta \,|\eta|^4 + \lambda_\rho \,|\rho|^4 + \lambda_\chi \,|\chi|^4 + \lambda_\sigma \,| \sigma|^4 \nn\\
&+& \lambda_{\eta \rho} \,|\eta|^2|\rho|^2+ \lambda_{\eta \chi}\, |\eta|^2|\chi|^2 +\lambda_{\eta \sigma} \,|\eta|^2 |\sigma|^2 + \lambda_{\rho \chi} \,|\rho|^2|\chi|^2+ \lambda_{\rho \sigma} \,|\rho|^2|\sigma|^2 +\lambda_{\chi \sigma} |\chi|^2|\sigma|^2 \nn\\
&+& \lambda_{\eta \rho 2}\,| \eta^\dagger\rho|^2+ \lambda_{\eta \chi 2}\,|\eta^\dagger \chi|^2+ \lambda_{\rho \chi 2}\,|\rho^\dagger \chi|^2 + (\lambda_4 (\sigma \rho [\chi \,\eta] ) + h.c.),
\eea 
with $\lambda_{\eta \rho 2},\lambda_{\eta \chi 2},\lambda_{\rho\chi 2}>0$, 
$M_\rho^2\gg w^2 \gg v^2\gg 0$ and
$\lambda_4<0$ is real after appropriate rephasing.

Upon substituting the field decompositions in Eqs. (\ref{sctripdec}) and (\ref{sigmadec}) into the potential above, the minimum conditions below follow 
\begin{eqnarray}\label{mincon}
\lambda_4u w v_\sigma   + v \left( \lambda_{\eta\rho}u^2 +2 \lambda_\eta v^2  + 
    \lambda_{\eta \chi} w^2   + \lambda_{\eta\sigma} v_\sigma^2  + 2 \mu_{\eta}^2\right)&=&0~, \\
    \lambda_4 v w v_\sigma  + u \left( \lambda_{\eta\rho} v^2 + 2 \lambda_\rho u^2  + 
    \lambda_{\rho \chi} w^2  + \lambda_{\rho\sigma} v_\sigma^2  + 2 M_{\rho}^2\right)&=&0~, \nn\\
     \lambda_4 u  v v_\sigma  +  w \left(\lambda_{\rho\chi} u^2 + 2 \lambda_\chi w^2  + \lambda_{\eta\chi} v^2 + 
    \lambda_{\chi\sigma} v_\sigma^2 + 2 \mu_{\chi}^2\right)&=&0~, \nn\\
    \lambda_4 u v w  + v_\sigma \left( \lambda_{\rho\sigma} u^2  +2 \lambda_\sigma v_\sigma^2  +  \lambda_{\eta\sigma} v^2 + \lambda_{\chi\sigma} w^2  + 2 \mu_{\sigma}^2\right)&=&0~,\nn
\end{eqnarray} 
which allow us to eliminate the quadratic mass parameters, $\mu_\eta^2$, $\mu_\chi^2$, $M_\rho^2$ and  $\mu_\sigma^2$, by writing them as functions of the vevs and dimensionless couplings.\,\footnote{%
We have checked that retaining a vev for $\chi_1^0$ and solving the minimization equations allow for a solution with the pattern in \eqref{sctripdec}.
}
These conditions also indicate that the coupling constants $\lambda_{t\sigma}$, with $t = \eta, \rho,\chi$, governing the interactions between the scalar singlet $\sigma$ with the other scalars,  are naturally suppressed.

\subsection{The spectrum} \label{ScSp}

We analyze now the quadratic terms of the potential to obtain the scalar particle spectrum. For the charged fields, we find that $\eta_2^+$ and $\rho_1^+$ mix and, after diagonalization, give rise to a physical charged scalar field, $\phi_1^+$, and a charged Goldstone boson, $G_1^+$. The physical fields are obtained from their relation with the symmetry states
\be 
\begin{pmatrix}
G_1^+ \\ \phi_1^+   
\end{pmatrix}
=
\begin{pmatrix}
\cos\theta_1 & \sin\theta_1 \\ - \sin\theta_1 & \cos \theta_1
\end{pmatrix}
\begin{pmatrix}
\eta_{2}^+ \\ \rho_1^+  
\end{pmatrix}~, ~~\mbox{with}~~ \tan(2\theta_1) = \frac{2 u v }{u^2 - v^2}
\approx -2\frac{u}{v}
~,
\ee 
and the mass of $\phi_1^+$ is 
\be \label{c1m}
m_1^2 =  \frac{u^2+v^2}{2} \left( \lambda_{\eta \rho 2}  - \frac{ \lambda_4 v_\sigma w }{u v}  \right)
\approx -\frac{\lambda_4v_\sigma w v}{2u}
~,
\ee
with $\lambda_4 <0$. This mass scale, which is much larger than $w$, is basically the effective mass scale of the triplet $\rho$ when it is integrated out; see \eqref{M:eff} and the discussion around it.
Similarly, $\chi_2^+$ and $\rho_3^+$ are mixed and can be written in terms of the independent states: $\phi_2^+$ and $G_2^+$, another charged scalar and Goldstone boson respectively, 
\be 
\begin{pmatrix}
G_2^+ \\ \phi_2^+   
\end{pmatrix}
=
\begin{pmatrix}
\cos\theta_2 & \sin\theta_2 \\ - \sin\theta_2 & \cos \theta_2
\end{pmatrix}
\begin{pmatrix}
\chi_{2}^+ \\ \rho_3^+  
\end{pmatrix}~, ~~\mbox{with}~~ \tan(2\theta_2) = \frac{2 u w}{u^2-w^2}
\approx -2\frac{u}{w}
~,
\ee 
where $\phi_2^+$ gets a mass given by
\be \label{c2m}
m_2^2 =  \frac{u^2+w^2}{2}\left( \lambda_{\rho \chi 2}  - \frac{ \lambda_4 v_\sigma v }{u w}  \right)
\approx 
m_1^2
+\ums{2}\lambda_{\rho\chi 2}w^2
~.
\ee 

When it comes to the neutral scalars, we have that $\chi_1^0$ and $\eta_3^0$ mix to form a Goldstone boson $G^0$ and a non-hermitian neutral field $\phi^0$ with normalized states written as 
\be 
\begin{pmatrix}
 G^0  \\ \phi^0  
\end{pmatrix}
=
\begin{pmatrix}
\cos\theta_0 & \sin\theta_0 \\ - \sin\theta_0 & \cos \theta_0
\end{pmatrix}
\begin{pmatrix}
\chi_{1}^0 \\ (\eta_{3}^0)^\dagger
\end{pmatrix}~, ~~\mbox{with}~~ \tan(2\theta_0) = \frac{2 v w }{v^2 - w^2}
\approx -2\frac{v}{w}
~.
\ee 
The mass of the non-hermitian neutral field $\phi^0$ is given by 
\be \label{nhm}
m_0^2 =  \frac{v^2+w^2}{2}\left( \lambda_{\eta \chi 2}  - \frac{ \lambda_4 u v_\sigma  }{v w}  \right)
\approx
\frac{u^2}{v^2} m_1^2
+\ums{2}\lambda_{\eta\chi 2}w^2
~.
\ee 
We can see that the mass of the neutral $\phi^0$ lies much lower than the masses of the charged scalars $\phi_{1,2}^+$, and it may be at the TeV scale.

Since the angles $\theta_i$ are all small due to the hierarchical vevs, the charged and the neutral non-hermitian physical scalar states are essentially given by the components in the triplets, i.e., $\phi_1^+\approx \rho_1^+$, $\phi_2^+\approx \rho_3^+$, $\phi^0\approx \eta_{3}^0$. 

Considering the pseudoscalars $A_\eta$, $A_\rho$, $A_\chi$ and $A_\sigma$ that mix with each other, we find upon diagonalizing their $4\times4$ mass matrix that only one independent field combination gets a mass after spontaneous symmetry breaking, 
\be \label{psm}
m_A^2 = -\lambda_4 \frac{ u^2 v^2 v_\sigma^2+u^2 v^2 w^2+u^2 w^2 v_\sigma^2+v^2 w^2 v_\sigma^2}{2 u v w v_\sigma}
\approx
m^2_1\Big(1+\frac{u^2}{v^2}\Big)
~,
\ee 
and the physical state associated with it is given by:
\be 
A = \frac{1}{\sqrt{1+(v_\sigma/u)^2+(v_\sigma/v)^2+(v_\sigma/w)^2}} \left[\left(\frac{v_\sigma}{u}\right) A_\rho + \left(\frac{v_\sigma}{v}\right)A_\eta +\left(\frac{v_\sigma}{w}\right)A_\chi+A_\sigma\right]~.
\ee 
Since $v_\sigma/u$ is the largest coefficient, the pseudoscalar in the particle spectrum is mostly composed of $A_\rho$. It is worth pointing out that in the limit $\lambda_4 \to 0$, the potential in Eq. (\ref{V}) displays an additional global symmetry which prevents the pseudoscalar $A$ to become massive. This symmetry in the absence of $\lambda_4$ would, however, be spontaneously broken by $u$ and, as a consequence, $A$ would become a massless Goldstone boson. Thus, because the vanishing of $\lambda_4$ is intimately related with the appearance of a new symmetry, we expect $|\lambda_4|$ to be naturally small. 

Although three pseudoscalars remain massless after the spontaneous symmetry breaking, one of them will get a tiny mass from non-perturbative effects as well as gravitational corrections. This field is the axion, $a$, the pseudo-Goldstone boson associated with the spontaneous breaking of the PQ symmetry, and it can be defined as
\be 
a = \frac{1}{\sqrt{v^2+ v_\sigma^2}}\left[ - v A_\eta + v_\sigma A_\sigma \right].
\ee 
As expected, the invisible axion is mostly made of the imaginary part of the scalar singlet $\sigma$.  Through non-perturbative QCD effects the axion field gets a potential and, consequently, a mass given by (see Ref.~\cite{Tanabashi:2018oca} and references therein)
\be 
m_a \simeq 5.7 \times \left(\frac{10^9~\text{GeV}}{f_a}\right)\text{meV}~ %= 1.2~\text{meV},
\label{maxion}
\ee 
with the axion decay constant $f_a\approx v_\sigma/2$ as defined above.  

The remaining pseudoscalars are Goldstone bosons which will be absorbed by the vector sector 
\bea 
G_{A1} &=& \frac{1}{\sqrt{u^2 + w^2}} \left[u A_\rho - w A_\chi \right]~,\nn \\
G_{A2} &=& c_{A2} \left[v v_\sigma^2 (u^2+w^2) A_\eta- u w^2 (v^2+v_\sigma^2) A_\rho-u^2 w (v^2+v_\sigma^2) A_\chi+v^2 v_\sigma (u^2+w^2) A_\sigma\right]~,\nn\\
&&\mbox{with}~~c_{A2} = \left\{(u^2+w^2)^2(v^2+v_\sigma^2)v^2 v_\sigma^2+(u^2+w^2)(v^2+v_\sigma^2)^2u^2 w^2\right\}^{-1/2}~.
\eea 

Finally, we look at the mixing amongst the real scalars $S_\eta, S_\rho, S_\chi$ and $S_\sigma$ from where the SM Higgs field should emerge. In the symmetry basis $(S_\eta, S_\rho, S_\chi, S_\sigma)$, the following squared mass matrix is generated
\be\label{mm}
M_S^2=
\begin{pmatrix}
 2 \lambda_\eta  v^2-\frac{u v_\sigma w \lambda_4}{2 v} & u v \lambda_{\eta \rho} +\frac{v_\sigma w \lambda_4}{2} & v w \lambda_{\eta \chi} +\frac{u v_\sigma \lambda_4}{2} & v v_\sigma \lambda_{\eta \sigma} +\frac{u w \lambda_4}{2} \\
 u v \lambda_{\eta \rho} +\frac{v_\sigma w \lambda_4}{2} & 2 \lambda_\rho  u^2-\frac{v v_\sigma w \lambda_4}{2 u} & u w \lambda_{\rho \chi} +\frac{v v_\sigma \lambda_4}{2} & u v_\sigma \lambda_{\rho \sigma} +\frac{v w \lambda_4}{2} \\
 v w \lambda_{\eta \chi} +\frac{u v_\sigma \lambda_4}{2} & u w \lambda_{\rho \chi} +\frac{v v_\sigma \lambda_4}{2} & 2 \lambda_\chi  w^2-\frac{u v v_\sigma \lambda_4}{2 w} & v_\sigma  w \lambda_{\chi \sigma} +\frac{u v \lambda_4}{2} \\
 v v_\sigma \lambda_{\eta \sigma} +\frac{u w \lambda_4}{2} & u v_\sigma \lambda_{\rho \sigma} +\frac{v w \lambda_4}{2} & v_\sigma w \lambda_{\chi \sigma} +\frac{u v \lambda_4}{2} & 2 \lambda_\sigma v_\sigma^2-\frac{u v w \lambda_4}{2 v_\sigma} 
\end{pmatrix}~.
\ee

As the diagonalization of the mass matrix above is clearly not as straightforward as in the previous cases, let us consider some simplifications. First, we remind ourselves that the couplings $\lambda_{t\sigma}\ll 1$, with $t=\eta,\rho,\chi$, following the relations in Eq. (\ref{mincon}). Second, as discussed below Eq. (\ref{psm}), $\lambda_4$ is expected to be small since in the limit that it goes to zero a new global symmetry shows up. These features, together with Eq. (\ref{mm}), tell us that the scalar $S_\sigma$ is effectively decoupled from the other fields, and its mass is proportional to $v_\sigma$. 

The SM symmetry breakdown is effectively governed by $v$ and, thus, we expect the main contribution to the Higgs boson, with a mass of $125$ GeV, to come from $S_\eta$. When taking $v_\sigma = 10^{10}$ GeV, $w = 10^4$ GeV, $v=246$ GeV and $u = 1$ GeV, as before, the correct Higgs mass can be obtained with, for example, $\lambda_4 = -10^{-4}$, $\lambda_\eta = 0.3641$, and the remaining coupling constants of the order of $10^{-1}$. In such an instance, the physical state is given by 	
\be 
h = -(0.999) S_\eta - (4 \times 10^{-3}) S_\rho + (3.7 \times 10^{-2}) S_\chi~,
\ee 
the absolutely dominant contribution coming from $S_\eta$, as expected. The other two massive scalar fields get the following masses:
\be 
m_{H_1} \simeq 4.47\times10^3~\mbox{GeV} ~~\mbox{and}~~m_{H_2} \simeq 1.1\times 10^6~\mbox{GeV}~,
\ee 
where $H_1$ is similarly dominated by $S_\chi$ and $H_2$ by $S_\rho$. For the sake of completeness, using this particular solution, we obtain the masses of the other scalars in the theory according to the Eqs. (\ref{c1m}), (\ref{c2m}), (\ref{nhm}) and (\ref{psm}).
The two charged scalars $\phi_{1,2}^{\pm}$ and the pseudoscalar $A$ get quite degenerate masses $m_1\simeq m_2 \simeq m_A \simeq 1.1 \times 10^6$ GeV as expected while we find $m_0 \simeq 5.0 \times 10^3$ GeV for 
the non-hermitian neutral scalar $\phi^0$.

Turning back to Eq.~(\ref{maxion}) if we take into account the value $v_\sigma=10^{10}$ GeV, used in the Sec.~\ref{ssec.doub} to exemplify the neutrino mass generation mechanism, the corresponding axion mass is $m_a\approx 1.1~\text{meV}$. It is still possible to have other values for this mass without affecting significantly the mass hierarchy pattern for the fermions in Sec.~\ref{sec.Fspec}. As an example, we could have $v_\sigma=10^{11}$ GeV, which implies $m_a\approx 1.1\times 10^{-4}~\text{eV}$, without modifying the entries of the neutrino mass matrix with a mild tuning of $\tilde{h}^S = 10^{-3}$ in Eq.~(\ref{Mnumass}) and keeping the same value for $\kappa$ (which requires just a rescaling of $\Lambda^\prime$). 

Axions with mass at the $\text{meV}$ scale could be the dominant component of cold dark matter of the Universe in post-inflationary PQ symmetry breaking scenarios as studied in Ref~\cite{Ringwald:2015dsf}. There it was shown that for some types of DFSZ models, which have domain wall number $N_{DW}=6$, an axion with a mass $m_a\approx(0.6-4)~\text{meV}$ can be a dominant component of cold dark matter. For the KSVZ models, which have $N_{DW}=1$, the axion mass would be $m_a\approx(0.8-1.3)\times 10^{-4}~\text{eV}$~\cite{Ringwald:2015dsf}. The model we are dealing with in this work is a sort of DFSZ-KSVZ hybrid model having $N_{DW}=2$ and a precise determination whether the mass in Eq.~(\ref{maxion}) allows the axion to account for all the dark matter in the universe, or a significant part of it, requires additional investigation. In any case we expect that the axion in this model can play the role of dark matter, once the model here allows for the axion mass in Eq.~(\ref{maxion}) to be in a relatively interesting range.

\subsection{On the integration of the heavy triplet and its effective vev}\label{rhoint}

In order to find the fermion spectrum, no operator containing $\rho$ explicitly was used in Sec.~\ref{sec.Fspec}. Instead, we have made use of effective operators with the form given by Eq. (\ref{effvev}). We want to show in this section that these operators emerge from the integration of the $\rho$, which is taken to be heavy compared to the other triplets. If we add $\rho$ and consider renormalizable operators only, instead of the operators in Eq. (\ref{D5}), we would have
\bea
-\mathcal{L}_\rho &= & y^\nu_{ij} [\overline{\Psi_{iL}}~\Psi_{jL}^c]~ \rho^*+y^e_{ij}
\overline{\Psi_{i L}} \,\rho\, e_{j R}
+y^d_{m} \overline{Q_{3 L}}\,\rho\, d_{m R}^\prime
+y^u_{an} \overline{Q_{a L}}\, \rho^*\, u_{n R}^\prime +h.c.~.
\label{Yukrho}
\eea
At low energies, meaning energies up to the $w=10^4$ GeV scale, the scalar singlet, whose vev is $v_\sigma \gg M_\rho, w, v$, can be effectively replaced in the potential by its vev. Furthermore, in the case that the triplet $\rho$ is much heavier than the other two, {\it i.e.}, $M_\rho \gg w, v$, it can be integrated out and its dominant contribution substituted back in the Lagrangian. From Eq. (\ref{V}), replacing $\sigma \to v_\sigma/\sqrt{2}$, $\rho$ can be integrated out, and its dominant low energy contribution will be
\be \label{effrho}
\rho = \frac{[\chi \,\eta]^*}{\Lambda} + \cdots~~~\mbox{with} ~~~\Lambda = \frac{\sqrt{2} M_{\rm eff}^2}{|\lambda_4| v_\sigma}~,
\ee 
where 
\eq{
\label{M:eff}
M_{\rm eff}^2 = M_\rho^2 + \lambda_{\rho\sigma} \frac{v_\sigma^2}{2}
+\dots
\,
}
is the $\rho$ effective mass, and the ellipsis represents the sub-dominant contributions which are neglected here. This result can be compared to Eq. (\ref{effvev}) where we defined the effective vev $u$ for the first time. The scale $\Lambda$ is therefore not a free parameter but a function of effective mass of the heavy triplet, the vevs of the scalar fields, and some of the dimensionless couplings present in the scalar potential of our model. It is now easy to see that when replacing $\rho$ as given by Eq. (\ref{effrho}) into Eq. (\ref{Yukrho}), we obtain Eq. (\ref{D5}).

Furthermore, we can use (\ref{effrho}) to derive constraints on some parameters of the model. As we take $u = 1$ GeV, we need that $\Lambda = 10^6$ GeV which, in turn, implies that $(M_{\rm eff}/\unit{GeV})\approx 10^3 \sqrt{|\lambda_4| (v_\sigma/\unit{GeV}) }$. If we assume again that $v_\sigma = 10^{10}$ GeV and $|\lambda_4| = 10^{-4}$, the following constraints follow
\be 
M_\rho \leq 10^6~\mbox{GeV} ~~~~\mbox{and}~~~~~ \lambda_{\rho\sigma} \leq 10^{-8}~. 
\ee 
As previously discussed, all couplings of the form $\lambda_{t\sigma}$, with $t$ varying amongst the triplets, are expected to be very suppressed, and the strong upper-bound on $\lambda_{\rho\sigma}$ found above only confirms that. Finally, the effective mass being of the order of $10^6$ GeV agrees with the masses found for the scalar fields associated with the $\rho$ triplet in the previous subsection and justifies the integration of $\rho$, since $M_{\rm eff}\gg w, v$.

\section{Flavor changing effects}
\label{Sec.FCNC}

Tree level flavor changing neutral currents (FCNCs) are well-known signatures of 3-3-1 models with generic $\beta$  \cite{Promberger:2007py,Promberger:2008xg,Buras:2012dp,Buras:2013dea,Buras:2016dxz,Queiroz:2016gif,Dong:2017ayu}, owing to the fact that one of the quark families transforms differently from the other two. More often than not, neutral currents mediated by the heavy $Z^\prime$ boson are the most relevant flavor changing interactions at tree level. However, other sources of FCNCs can also be present, for example, those mediated by the $Z$ boson through its small mixing with the $Z^\prime$, as seen in Eq. (\ref{ZZmix}), or those mediated by scalar fields.

The version considered here with $\beta = - 1 /\sqrt{3}$ (as well as that with $\beta = + 1 /\sqrt{3}$) is yet a more abundant source of FCNCs for two reasons. Firstly, it contains new quarks with electric charges of $-1/3$ and $2/3$, inducing new flavor changing contributions through their mixing with the SM quarks. Secondly, in addition to $Z$ and $Z^\prime$, the gauge spectrum presents another neutral field, $V^0$. 

Similar to the case with $\beta=1/\sqrt{3}$ studied in Ref.~\cite{Barreto:2017xix}, the present scalar sector contains only one scalar doublet around or below the TeV scale so that the only FCNC mediated by scalars at tree level is due to the small mixing between the heavy and the SM quarks. Such a mixing however, being proportional to several powers of the suppression factor $\kappa$, as discussed in Sec. \ref{sec.Fspec},  can be safely disregarded. Concerning the FCNCs mediated by the gauge bosons, we have again a similar situation to that in Ref.~\cite{Barreto:2017xix}. Due to the hierarchy among the different mass scales: $\kappa v, u, v$ and $w$, the expected flavor changing effects are in good agreement with current experimental constraints, such as those coming from meson mass differences. A thorough analysis on the effects of FCNCs to investigate, for instance, anomalies in B-physics is worth the attention in a future work. 

Given that our PQ symmetry with charges in Table~\ref{TPQ} is not flavor universal, constraints coming from meson decays emitting axions may be relevant\,\cite{Bjorkeroth:2018dzu}.
Considering our PQ scale of $v_\sigma\sim 10^{10}\,\unit{GeV}$ and the quark mass matrix structure in Eq. \eqref{quark:mass:pattern}, we expect a $s\to d$ transition to occur with axion emission with strength proportional to $|V^d_{sd}|\approx |4U^{d*}_{L3s}U^d_{L3d}|\sim  
|4V^{\rm CKM}_{ts}V^{\rm CKM}_{td}|
\sim 10^{-3}$, where we only retained the dominant contribution of left-handed $d$-type quarks.
This means that we can evade current bounds\,\cite{Bjorkeroth:2018dzu} but future bounds can constrain the naturality of our mass matrix structure.

\section{Conclusions}\label{conclusion}

In this work we have investigated a version of the 3-3-1 model defined by $\beta=-1/\sqrt{3}$ and augmented by an additional $\ZZ_9$ symmetry. This scenario leads to an accidental Peccei-Quinn symmetry which is spontaneously broken at a high energy scale by the vev of a gauge-singlet scalar field, giving rise to an invisible axion, which can play the role of dark matter, as well as allowing for a solution to the strong CP problem via the PQ mechanism. The 3-3-1 gauge symmetry, on the other hand, is broken effectively in two steps by the vevs of two $SU(3)_L$ scalar triplets that transform identically under the 3-3-1 gauge symmetry. The first breaking takes place at the TeV scale, and the second occurs at the electroweak symmetry breaking scale. A third scalar triplet is however required to break a residual symmetry that prevents some of the fermions to become massive. The scalar spectrum, up to the TeV scale, is compact and presents the interesting feature of being completely composed of neutral fields: a light axion, the SM Higgs boson, and three TeV scale neutral bosons. The remaining scalars can become very heavy ($\gg 10^3$ TeV) and are therefore not expected to be observed by any near-future experiments. In the gauge boson sector, in addition to the SM vector bosons, the following heavy fields are present: $V^0,\,(V^0)^\dagger,\,U^\pm$ and $Z^\prime$. The mixing between $Z^\prime$ and $Z$, the SM neutral gauge boson, for $Z^\prime$ masses around the TeV scale has been shown to be very suppressed $\varphi \simeq 10^{-4}$.  

The quark sector of the model contains a new up-type and two new down-type quarks, all of which get masses around the TeV scale. When considering the SM quarks, we have seen that due to the PQ symmetry and its associated scalar singlet $\sigma$, a suppression mechanism takes place leading to natural mass hierarchies between the third and the other two families. In the leptonic sector we have introduced six new neutral fields, three of which are gauge singlets. The Yukawa couplings required to describe the charged lepton masses are less suppressed than in the SM, and neutrinos become massive through the double seesaw mechanism whose implementation relies also on the PQ symmetry. Three (sub-eV) active, three GeV and three super-heavy ($\sim 10^8$ GeV) neutrinos make up the neutrino particle spectrum. As for the GeV neutrinos, we have shown that their mixing with the active neutrinos is small enough to evade experimental constraints, such as the one coming from the radiative muon decay $\mu \to e \gamma$ or neutrinoless double beta decay, but large enough to allow them to decay sufficiently earlier than the BBN epoch.

A common feature in 3-3-1 models like ours, built with one quark family transforming differently from the other two under the gauge symmetries, is the presence of tree-level FCNCs. In the present case however the flavor changing effects are well within the experimental limits due to the hierarchy among the different vevs as well as the suppressed interactions as a result of the PQ symmetry. 

Finally, we would like to mention that the model has a potentially interesting phenomenology involving the new particles, such as the scalars $H_1$ and $\phi^0$, at the TeV scale. For example, the production of $H_1$ through gluon fusion through the new quarks might furnish distinct signals such as the diphoton decay $gg\rightarrow H_1\rightarrow \gamma\gamma$, and a  pair of Higgs bosons $gg\rightarrow H_1\rightarrow hh$.  

\acknowledgments
This research was partially supported by the Conselho Nacional de Desenvolvimento Cient\'{\i}fico e Tecnol\'ogico (CNPq), by grants 306636/2016-6 (A.G.D.) and
308578/2016-3 (C.C.N.).
Financial support by Funda\c{c}\~{a}o de Amparo \`{a} Pesquisa do Estado de S\~ao Paulo (FAPESP) is also acknowledged under the grants 2014/19164-6 (C.C.N.) and 2017/23027-2 (J.L.). This study was financed in part by the Coordena\c{c}\~{a}o de Aperfei\c{c}oamento de Pessoal de N\'ivel Superior - Brasil (CAPES) - Finance Code 001 (D.D.L.).

%%%%%%%%%%%%%%%%%%%%%%%%
\end{document}